\documentclass{emulateapj}

\def\gsim{\mathrel{\rlap{\lower 4pt \hbox{\hskip 1pt $\sim$}}\raise 1pt
\hbox {$>$}}}
\def\lsim{\mathrel{\rlap{\lower 4pt \hbox{\hskip 1pt $\sim$}}\raise 1pt
\hbox {$<$}}}

\begin{document}

\title{Type IIb Supernova 2013df Entering Into An Interaction Phase: \\
A Link between the Progenitor and the Mass Loss$^{*}$
}  

\author{
K. Maeda\altaffilmark{1,2}, T. Hattori\altaffilmark{3}, D. Milisavljevic\altaffilmark{4}, G. Folatelli\altaffilmark{5,2}, M.R. Drout\altaffilmark{4}, H. Kuncarayakti\altaffilmark{6,7}, \\ 
R. Margutti\altaffilmark{4}, A. Kamble\altaffilmark{4}, A. Soderberg\altaffilmark{4}, M. Tanaka\altaffilmark{8}, M. Kawabata\altaffilmark{9}, K.S. Kawabata\altaffilmark{9,10}, \\
M. Yamanaka\altaffilmark{11,12}, K. Nomoto\altaffilmark{2,**}, J.H. Kim\altaffilmark{3}, J.D. Simon\altaffilmark{13}, M.M. Phillips\altaffilmark{14}, J. Parrent\altaffilmark{4}, \\ 
T. Nakaoka\altaffilmark{9}, T.J. Moriya\altaffilmark{15}, A. Suzuki\altaffilmark{1}, K., Takaki\altaffilmark{9}, M. Ishigaki\altaffilmark{2}, I. Sakon\altaffilmark{16}, \\ 
A. Tajitsu\altaffilmark{3}, M. Iye\altaffilmark{8}
}
\altaffiltext{*}{Based on data collected at the Subaru Telescope operated by the National Astronomical Observatory of Japan, and also at the MMT Observatory, a joint facility of the Smithsonian Institution and the University of Arizona.}
\altaffiltext{1}{Department of Astronomy, Kyoto University
Kitashirakawa-Oiwake-cho, Sakyo-ku, Kyoto 606-8502, Japan; keiichi.maeda@kusastro.kyoto-u.ac.jp .}
\altaffiltext{2}{Kavli Institute for the Physics and Mathematics of the 
Universe (WPI), University of Tokyo, 
5-1-5 Kashiwanoha, Kashiwa, Chiba 277-8583, Japan}
\altaffiltext{3}{Subaru Telescope, National Astronomical Observatory of Japan, 650 North A'ohoku Place, Hilo, Hawaii 96720} 
\altaffiltext{4}{Harvard-Smithsonian Center for Astrophysics, 60 Garden Street, Cambridge, MA 02138}
\altaffiltext{5}{Instituto de Astrof\'isica de La Plata, CONICET, Paseo del Bosque S/N, B1900FWA, La Plata, Argentina}
\altaffiltext{6}{Millennium Institute of Astrophysics, Casilla 36-D, Santiago, Chile}
\altaffiltext{7}{Departamento de Astronom\'ia, Universidad de Chile, Casilla 36-D, Santiago, Chile}
\altaffiltext{8}{National Astronomical Observatory of Japan, Mitaka, Tokyo 181-8588, Japan} 
\altaffiltext{9}{Department of Physical Science, Hiroshima University, Kagamiyama, Higashi-Hiroshima 739-8526, Japan} 
\altaffiltext{10}{Hiroshima Astrophysical Science Center, Hiroshima University, Kagamiyama, Higashi-Hiroshima, Hiroshima 739-8526, Japan} 
\altaffiltext{11}{Department of Physics, Faculty of Science and Engineering, Konan University, Okamoto, Kobe, Hyogo 658-8501, Japan}
\altaffiltext{12}{Kwasan Observatory, Kyoto University, 17-1 Kitakazanohmine-cho, Yamashina-ku, Kyoto, 607-8471}
\altaffiltext{13}{Observatories of the Carnegie Institution for Science, 813 Santa Barbara St, Pasadena, Ca 91101}
\altaffiltext{14}{Carnegie Observatories, Las Campanas Observatory, Casilla 601, La Serena, Chile}
\altaffiltext{15}{Argelander Institute for Astronomy, University of Bonn, Auf dem H\"ugel 71, 53121, Bonn, Germany}
\altaffiltext{16}{Department of Astronomy, Graduate School of Science, The University of Tokyo, Hongo 7-3-1, Bunkyo-ku, Tokyo, 113-0033}
\altaffiltext{**}{Hamamatsu Professor}

\begin{abstract}
We report the late-time evolution of Type IIb Supernova (SN IIb) 2013df. SN 2013df showed a dramatic change in its spectral features at $\sim 1$ year after the explosion. Early on it showed typical characteristics shared by SNe IIb/Ib/Ic dominated by metal emission lines, while later on it was dominated by broad and flat-topped H$\alpha$ and He I emissions. The late-time spectra are strikingly similar to SN IIb 1993J, which is the only previous example clearly showing the same transition. This late-time evolution is fully explained by a change in the energy input from the $^{56}$Co decay to the interaction between the SN ejecta and dense circumstellar matter (CSM). The mass loss rate is derived to be $\sim (5.4 \pm 3.2) \times 10^{-5} M_{\odot}$ yr$^{-1}$ (for the wind velocity of $\sim 20$ km s$^{-1}$), similar to SN 1993J but larger than SN IIb 2011dh by an order of magnitude. The striking similarity between SNe 2013df and 1993J in the (candidate) progenitors and the CSM environments, and the contrast in these natures to SN 2011dh, infer that there is a link between the natures of the progenitor and the mass loss: SNe IIb with a more extended progenitor have experienced a much stronger mass loss in the final centuries toward the explosion. It might indicate that SNe IIb from a more extended progenitor are the explosions during a strong binary interaction phase, while those from a less extended progenitor have a delay between the strong binary interaction and the explosion. 
\end{abstract}


\section{Introduction}

Evolution of a massive star in the final stage toward the supernova (SN) explosion is one of the main unresolved issues in modern stellar astrophysics. Evolutionary paths to SNe IIb/Ib/Ic, called stripped-envelope (SE-) SNe, have been actively debated -- it is widely accepted that the progenitors of SE-SNe have lost most (SNe IIb) or nearly all (SNe Ib/c) of an H-rich envelope \citep{filippenko1997}, and the unresolved mass loss mechanism should be a direct consequence of the progenitor evolution \citep{nomoto1993,woosley1994,georgy2012,benvenuto2013}. SNe IIb are in particular important objects in this respect. They have a little but non-negligible amount of H-rich envelope ($\lsim 1 M_{\odot}$) present at the time of the explosion \citep{shigeyama1994,bersten2012}, representing a link between SNe II (with a red supergiant progenitor) and SNe Ib/c (with a Wolf-Rayet like progenitor). Direct progenitor search in deep pre-explosion images points to a diversity in the progenitor, highlighted by different extension of the progenitor envelope for SNe IIb 1993J \citep{maund2004} and 2011dh \citep{vandyk2013}; while both were unprecedentedly identified as Yellow Supergiant (YSG) stars, their sizes were significantly different with $R \sim 600 R_{\odot}$ for the former and $\sim 200 R_{\odot}$ for the latter. 

SN IIb 2013df was discovered on June 7.9, 2013 UT soon after the explosion \citep{chiabattari2013,cenko2013} in NGC4414 at $\sim 16.6$ Mpc \citep{freedman2001}. A progenitor candidate has been identified in pre-SN HST images, whose nature is quite similar to the progenitor of SN 1993J, interpreted as a giant with a radius of $545 \pm 65 R_{\odot}$ \citep{vandyk2014}. Properties as an SN are also similar to those of SN 1993J \citep{morales2014,vandyk2014,benami2015}; both showed similarly a bright `early-cooling' phase, and similar `peak' light curves (LCs) and spectral evolutions around the maximum light. Throughout this paper, we refer the emission from the adiabatically expanding SN ejecta following the shock breakout as the `early-cooling emission'; the strength and duration of this emission are sensitive to the progenitor radius \citep[e.g., ][]{nakar2014}, and it lasts for less than $\sim 10$ days after the shock breakout for SNe IIb. The phase is followed by the emission powered by radioactive decays of $^{56}$Ni to $^{56}$Co and then to $^{56}$Fe, corresponding to the main part of the LC which is peaked at $\sim 20$ days after the breakout for SNe IIb. This `peak' LC and spectral properties are mostly determined by the properties of the SN ejecta rather than those of the progenitor. While the ejecta mass has been estimated slightly smaller in SN 2013df \citep{morales2014}, these two SNe seem to be almost a twin in terms of the progenitor and the SN properties. A contrast of these SNe to another well-studied SN IIb 2011dh is interesting -- while showing the similar peak LCs and spectral properties (thus similar SN properties), its progenitor was less extended \citep{vandyk2013} as was consistent with the faint early-phase cooling emission \citep{bersten2012}. 

In this paper, we report the late-time optical evolution of SN 2013df up to $\gsim 600$ days after the explosion. \S 2 presents observations and data reduction. \S 3 shows that the spectral and light curve evolutions are found to be dramatically changed in the late phase, showing clear signatures that the SN is now powered by the interaction between the SN ejecta and circumstellar matter (CSM). Among SE-SNe, such a transition was observed previously only for SN 1993J, and SN 2013df is the second example.  Indeed, the present study reveals that SN 2013df is a twin of SN 1993J, not only in the early phase but also in the late phase. This again provides a striking difference to SN 2011dh, for which a strong interaction was not observed in the late-phase. In \S 4, we analyze the late-time spectra and light curves, deriving properties of CSM, and discuss a relation between the natures of a progenitor and the mass loss. The paper is closed in \S 5. 

\section{Observations and Data Reduction}

\subsection{Spectra}
Our spectroscopic observations were performed on MJD 56648.6 (22 December 2013; day 199) and on MJD 57075.4 (22 February 2015; day 626) with the 8.2m Subaru telescope equipped with the Faint Object Camera and Spectrograph \citep[FOCAS; ][]{kashikawa2002}. Additional two spectra were obtained on MJD 56604.5 (8 November 2013; day 155) and on MJD 56783.3 (6 May 2014; day 333) with the 6.5m MMT telescope using the Blue Channel instrument \citep{schmidt1989}. The dates indicated here are measured from the well-defined explosion date \citep[MJD 56449.5:][]{morales2014,vandyk2014}, and this definition is used throughout the paper. 

The setup for the Subaru/FOCAS observations is the following: We used 0.8" slit and the B300 and R300 grisms on day 199 covering 3,650--10,000\AA, and 0.8" offset slit and the B300 grism on day 626 covering 4,700--9,000\AA. The slit direction is set at PA = 0 using an Atmospheric Dispersion Corrector (ADC). The spectral resolution is $\sim 500$. The exposure times are 600 s (in AB) and 3600 s (in ABBA), respectively.  For flux calibration and telluric absorption correction, Feige 34 \citep{oke1990} was observed in both nights. For MMT Blue Channel observations we used a 1" slit and the 300 line mm$^{-1}$ grating covering 3,200-8,500\AA. Both observations were at the parallactic angle. The spectral resolution is $R = 740$. The exposure times are 600 s (single) and $2 \times 1200$ s, respectively. For flux calibration and telluric absorption correction, standard stars from \citet{oke1990} and \citet{massey1990} were observed.

The spectra are reduced following standard procedures with {\em IRAF/PyRAF},\footnote{IRAF is distributed by the National Optical Astronomy Observatory, which is operated by the Association of Universities for Research in Astronomy, Inc., under cooperative agreement with the National Science Foundation. PyRAF is a product of the Space Telescope Science Institute, which is operated by AURA for NASA.} including bias subtraction, flat fielding, cosmic ray rejection \cite[with LAcosmic; ][]{vandokkum2001}, sky subtraction, 1D spectral extraction, wavelength calibration using ThAr or HeNeAr lamps, and flux calibration. The wavelength solution was checked with the atmospheric sky emission lines, and when necessary a small shift was further applied. 

\subsection{Imaging and Photometry}

For photometry, we obtained $V$ (20 s) and $R$ (20 s) images on day 199 and $V$ (120 s) and $R$ (180 s) images on day 626 with Subaru/FOCAS. On day 199, the standard stars around PG1525-071 \citep{landolt1992} were observed. 
We also obtained the SDSS $r$-band images at three epochs (300 s for each), on day 262, 298, and 348, using MMTCam. 

PSF photometry was performed on day 199 using the images around the PG1525-071 under photometric condition. This results in $V = 19.2 \pm 0.1$ mag and $R = 18.5 \pm 0.1$ mag. The photometry on the MMTCam images results in $r = 19.75 \pm 0.10$ (day 262), $20.08 \pm 0.08$ (day 298), and $20.59 \pm 0.09$ (day 348). The MMTCam images were calibrated using observations of SDSS field stars. Since the SN was found to be still relatively bright in these epochs, we have not subtracted a reference image for photometry with these images. 

The SN has faded substantially on day 626, but it was clearly detected in both $V$ and $R$ bands (Fig. 1). Since the SN magnitude is now comparable to the background which shows substantial fluctuation around the SN site, the photometry on day 626 requires a careful analysis. First we performed relative photometry between the images on day 199 and 626, then we subtracted the pre-HST WFPC2/F555W image on 29 April 1999 \citep[Obs ID: 8400, PI: Keith Noll; see also][]{vandyk2014} from our $V$-band image on day 626. Thereby we have obtained $V \sim 22.6$ mag. This was used to anchor the flux scale in the spectrum, then the $R$ and $I$-band magnitudes were derived as $R \sim 22.0$ mag and $I \sim 22.5$ mag from spectroscopic photometry. Details on the photometry will be presented in a separate paper (Maeda et al., in prep.). In any case, the uncertainty of the flux level even by a factor of $\sim 2$ would not affect our analyses and conclusions in this paper. 

\section{Results}

Figure 2 shows the time sequence of late-time spectra of SN 2013df from day 155 to day 626. Figure 3a shows the spectra of SN 2013df at day 199 and day 626, as compared to those of SNe IIb 1993J \citep{matheson2000a,matheson2000b} and 2011dh \citep{shivvers2013} at similar epochs. The comparison spectra are downloaded from The Weizmann interactive supernova data repository \citep{ofer2012}.\footnote{http://wiserep.weizmann.ac.il.} As discussed by \citet{morales2014}, the spectra of SN 2013df at $\lsim 200$ days since the explosion are overall typical for SNe IIb/Ib/Ic, dominated mostly by forbidden lines of metals and some allowed transitions. There are two especially interesting features: (1) The ratio of [OI] 6300, 6363 to [Ca II] 7291, 7324 is low, and (2) there is an excessive emission on the red side of the [O I]. The feature (1) can be interpreted as a relatively low-mass progenitor \citep{fransson1989,maeda2007a}, indicating the importance of the binary interaction to get rid of most of the hydrogen envelope before the explosion to become an SN IIb. The feature (2) is quite common in SNe IIb, as exemplified by SNe 1993J \citep{patat1995,matheson2000a,matheson2000b} and 2011dh \citep{sahu2013,shivvers2013}. The similar feature is also seen in some SNe Ib/c \citep{sollerman1998,maeda2007b,taubenberger2011,folatelli2014a}. A straightforward identification is H$\alpha$, but the solid identification is difficult -- the feature could be attributed to [N II] 6548,6583 at $\sim 200$ days \citep{jerkstrand2015}. 

A dramatic change was observed in the spectrum taken on day 626. There is little trace of metal emission lines anymore, but the spectrum is characterized by broad emission features including those around $\sim 6,550$\AA\ and $\sim 5,800$\AA. Indeed, a resemblance of this spectrum to that of SN 1993J at a similar epoch is very striking (Figure 3b). At such a late epoch, a canonical $^{56}$Co-powered nebular emission model is unable to explain these broad emission lines \citep{jerkstrand2015}, and there is almost no doubt that these features are H$\alpha$ and He I 5876/Na ID (plus [N II] on the blue side), respectively. This spectrum shows that the interaction between the SN ejecta and CSM is now a predominant power source. The spectrum on day 333 nicely shows the development of H$\alpha$ and the onset of a transition from the $^{56}$Co-powered emission to the SN-CSM interaction-powered emission (Figure 2). This is the second example, with SN 1993J as the first example \citep{matheson2000a,matheson2000b}, where an SN IIb/Ib/Ic shows such a transition. A marginal detection of a broad H$\alpha$ was also reported for SN 2011dh at a similar epoch \citep{shivvers2013}, but the strength of the H$\alpha$ does not compare to those found in SNe 2013df and 1993J (Figure 3a). 

Figure 4a compares the profiles of H$\alpha$ emission seen in SNe 2013df and 1993J at $\gsim 600$ days. First, for both SNe 2013df and 1993J, the H$\alpha$ shows a boxy profile with a steep cut-off on both sides. This indicates that the main part of the emission is originated in a thin shell with the velocity of $\sim 10,000$ km s$^{-1}$ (see the model curves in Figure 4a). There is however noticeable difference in details, where SN 2013df shows an excessive emission near the rest wavelength but blueward, while SN 1993J shows a suppression in the same wavelength region. A similar behavior is marginally visible in both blue and red edges, where SN 2013df shows an excess \citep[see also][]{morales2014}. The difference in the line profiles is probably caused by a different structure and/or viewing direction (see below for further support). If it is put into a unified context, a density enhancement in the equatorial direction, with SN 2013df viewed from the polar direction and SN 1993J from the equatorial direction, could explain the main difference seen around the rest wavelength. There are another couple of possible, but less likely, explanations for this relatively narrow, central component in SN 2013df: (1) This might reflect an unshocked CSM ionized by the high-energy radiation from the shocked region. The velocity is however too high ($\sim 2,000$ km s$^{-1}$) . (2) This might reflect swept-up materials from a companion star \citep{hirai2014}, analogous to what has been searched for in SNe Ia \citep{lundqvist2013,maeda2014b}. If we apply the model prediction for SNe Ia, the feature in SN 2013df seems too strong to be interpreted this way. 

Figure 4b shows the feature at $\sim 5,800$\AA. The red part of the feature matches well to H$\alpha$ assuming this feature is either He I 5876 or Na ID. This supports an idea that this feature is also powered by the SN-CSM interaction. The blue part shows some excessive emissions, which might be contaminated by [N II] and [O I], either from the SN ejecta, the interacting region, or the unshocked but irradiated CSM. 

Figures 4c shows the evolution of SN 2013df from day 155 to day 626. If we scale the spectrum at day 626 multiplied by a factor of $\sim 20$, $8$, and $2.5$, the emission at the red side of the [O I] as seen in the spectra at day 155, 199, and 333 matches to the H$\alpha$ profile very well. This suggests that this feature seen at as early as day 155 was already dominated by H$\alpha$. This was speculated by \citet{morales2014}, but the late-time spectrum is necessary for the conclusive identification, given possible alternative explanations \citep{jerkstrand2015}. Since the profile of H$\alpha$ does not show substantial evolution from day 155 to day 626 (except for the high-velocity edge; see below and \S 4.1), the line profile should reflect a geometrical nature of the emitting region, rather than the contributions by other lines, e.g., [N II]. 

Figure 4d compares the red part of the H$\alpha$ profile on day 199 and day 626. While the overall profile is kept nearly the same, there is a decrease in the velocity corresponding to the wavelength at the cut-off, namely the velocity of the emitting shell. This is qualitatively in agreement with the expectation from the SN-CSM interaction, where the shell created at the interacting region should be decelerated as a function of time. This will be further discussed in \S 4.1.  

Figure 5 shows long-term $V$ and $R$-band light curves of SN 2013df as compared to those of SNe 1993J and 2011dh. In this figure, we have converted the SDSS $r$-band magnitudes obtained by the MMTCam to the $R$-band magnitudes assuming a constant offset of $-0.15$ mag \citep{morales2014}. Up to $\sim 200$ days, the three SNe show mutually similar evolutions, suggesting that the ejecta mass ($M_{\rm ej}$) and the kinetic energy ($E_{\rm K}$), or specifically the optical depth to $\gamma$-rays from $^{56}$Co as scaled by $M_{\rm ej}^2/E_{\rm K}$ \citep{maeda2003}, are similar; in this sense, the ejecta mass obtained by \citet{morales2014}, $M_{\rm ej} = 1.2 \pm 0.2 M_{\odot}$ and $E_{\rm K} = 0.8 \pm 0.4 \times 10^{51}$ erg for SN 2013df, as compared to $M_{\rm ej} = 2.7 \pm 0.8 M_{\odot}$ and $E_{\rm K} = 1.2 \pm 0.2 \times 10^{51}$ erg for SN 1993J \citep{young1995}, is probably underestimated. On day 626, SN 2013df turns out to be brighter than SN 2011dh by $\sim 2$ mag in the $V$-band and $\sim 2.5$ mag in the $R$-band (when normalized at their peak magnitudes), as is consistent with the power source dominated by the SN-CSM interaction. This suggests that the main energy source had changed at $\sim 1$ years after the explosion. Indeed, the $R$-band light curves of SNe 2013df and 1993J both deviate from that of 2011dh at $\sim 200 - 300$ days (as is also seen in the $V$-band light curve of SN 1993J), exceeding the magnitudes of SN 2011dh at the corresponding epochs. Together with the spectral similarity, this also suggests that the SN-CSM interaction property is very similar between SN 2013df and 1993J. We note that the similarity in the SN-CSM interaction is also supported by an independent analysis of radio and X-ray properties of SN 2013df \citep[][]{kamble2015} as further mentioned in \S 4. 

Currently SN 2013df is much brighter in the $V$ and $R$-bands than the expectation from the $^{56}$Co-decay heating, and this would raise a hurdle for searching for disappearance of the progenitor candidate and/or for emergence of a possible blue companion star \citep{maund2004,vandyk2013,folatelli2014b,fox2014}. \citet{vandyk2014} identified the candidate progenitor as a giant with $R_{\rm eff} = 545 \pm 65 R_{\odot}$ with $V = 24.5$ mag and $I = 23.1$ mag ($R \sim 24$ mag if simply interpolated). If there would have been only the $^{56}$Co decay power (following the light curve of SN 2011dh), the $V$ and $R$-band magnitudes should have been already comparable to or even fainter than those of the progenitor candidates on day 626. Due to the additional power input by the interaction, the epoch when the SN becomes fainter than the progenitor candidate should be delayed. 

In Figure 5b, we show the contribution from the interaction power in the $R$-band light curves of SN 2013df (denoted by star symbols) estimated as follow: Assuming that the H$\alpha$ luminosity follows the interaction power, we estimated the interaction power at the epochs where the spectra are available (Figure 4c). Then we assumed that the $R$-band magnitude at day 626 is dominated by the interaction, and normalized the interaction power at this epoch. The same evolution is assumed for the $V$-band. Avoiding possible overestimate of the interaction power in the relatively early phase (i.e., day 155, through the contribution by [N II] and continuum), we fit the interaction power thus estimated on days 199, 333, and 626. With only three points the fit is uncertain, but we should be able to provide some meaningful estimate on when the confirmation of the progenitor is possible. If we assume an exponential decline in luminosity then the fit results in $L \propto e^{-t/210 \ {\rm days}}$, while if we assume a power law then $L \propto t^{-1.8}$. In the exponential case, the SN will become fainter than the progenitor candidate at $\sim 1,000$ days after the explosion both in $V$ and $R$-bands, i.e., February - March 2016. In the case of the power law decline, it will further be delayed to $\sim 1,600$ days after the explosion, i.e., October - November 2017. We suggest continues follow-up observations of SN 2013df to refine the prediction. This point will be further discussed in a separate paper (Maeda et al., in prep.).

\section{Discussion}

\subsection{Analyses of the H$\alpha$ and CSM around SN 2013df}

The observed broad and flat-topped H$\alpha$ emission is naturally expected as arising from the SN-CSM interaction. The interaction creates a reverse shock (RS), contact discontinuity (CD), and forward shock (FS) \citep{chevalier1982a}. The H$\alpha$ emission can be originated in the unshocked SN ejecta irradiated by X-ray photons created at the RS: Ionizations induced by high-energy radiations are followed by recombinations. This is also an interpretation for H$\alpha$ emission observed in middle-aged SNe II \citep{milisavljevic2012}. Alternatively, it could be contributed by the RS region if the temperature in the RS region is decreased to $\sim 10,000$ K. The latter situation is less likely as the RS is typically considered to have much higher temperature to emit X-rays for SNe IIb/Ib/Ic \citep[e.g., ][]{maeda2014a}, and we note that the FS region would have even much higher temperature. Indeed, analysis of the X-ray emission from SN 2013df does indicate the temperatures in the RS and FS regions as $\gsim 10^7$K and $\gsim 5 \times 10^{8}$K, respectively \citep{kamble2015}. In either case of the unshocked ejecta or the RS region, the outer edge of the H$\alpha$-emitting region can be identified as the CD. The emitting region may well be identified as nearly a whole H-rich envelope in the unshocked ejecta scenario while as a part of the region in the RS scenario. 

For the distance of $16.6 \pm 0.4$ Mpc, the H$\alpha$ luminosity is $\sim 8.8 \times 10^{38}$ erg s$^{-1}$ on day 199 and $\sim 1.1 \times 10^{38}$ erg s$^{-1}$ on day 626, with $E(B-V) = 0.1$ mag and $R_{V} = 3.1$. This indicates that the H$\alpha$ luminosity is similar to, or a bit stronger than, SN 1993J, for which it was derived to be $\sim 1.26 \times 10^{38}$ erg s$^{-1}$ on day 367 \citep{patat1995}. At day 626, the velocity of the shell is $\sim 10,000$ km s$^{-1}$ (Figures 4a and 4d). An uncertainty of $\sim 5$\% in the shell velocity is mainly from the uncertainty in defining the continuum. From a comparison between the H$\alpha$ profiles between day 199 and day 626 (Fig. 4d), we observe the velocity has decreased by $\sim 5 - 10$\% (where the main uncertainty is from the estimate of the shell velocity as mentioned above). The emitting region should be confined within a thin shell with $\Delta R_{\rm sh} / R_{\rm sh} \sim 0.2$ as constrained by the line profile (Figure 4a). Within the fitting error, $\Delta R_{\rm sh} / R_{\rm sh} \sim 0.1$ is acceptable, but $\Delta R_{\rm sh} / R_{\rm sh} \sim 0.3$ is rejected. 

As for the `He I' feature, the flux in the He I/Na ID at day 626 is by a factor of $\sim 3.2$ smaller than H$\alpha$ (excluding the excessive emission blueward: Figure 4b). Attributing this feature to He I 5876, we can derive the abundance in the emitting region. Assuming the Case B recombination both for H$\alpha$ and He I, the relative flux will be $L ({\rm H}_{\alpha}$)/$L({\rm He I})$ $\sim 1.3 n({\rm H}^{+})/n({\rm He}^{+})$ for a wide range of temperature (5,000 - 20,000K) where $n$'s are number densities \citep{osterbrock}. For the observed ratio of $3.2$, it requires $n({\rm H}^{+})/n({\rm He}^{+}) \sim 2.5$. Assuming that all the hydrogen and helium are singly ionized, the emitting region must be composed of $\sim 70$\% of H and $\sim 30$\% of He in number, or $\sim 40$\% of H and $\sim 60$\% of He in mass. Typical models for SN IIb adopt the He-rich H/He envelope with the mass fraction of He larger than that of H \citep[e.g., $Y \sim 0.8$ for  SN 1993J: ][]{shigeyama1994}, and our derived abundance in the emitting region is consistent with the expectation. 

\citet{patat1995} constrained the mass of the envelope hydrogen for SN 1993J on day 367. Since the derived geometry of the emitting region for SN 2013df is similar to the situation they considered, we use their equation 2 to estimate the envelope hydrogen mass for SN 2013df. In doing this, we have an additional constraint for the electron density as $1.4 n_{\rm H}^{+}$ (see above). Adopting the velocity at the CD as $V_{\rm sh} = 10,000$ km s$^{-1}$ and the luminosity at day 626, we derive $M({\rm H}) \sim 0.25 M_{\odot}$ at day 626 for the temperature of $10,000$K. Adopting $V_{\rm sh} = 11,000$ km s$^{-1}$ (see above; Figure 4d), we derive $M({\rm H}) \sim 0.14 M_{\odot}$ at day 199. We note these two estimates are consistent within the uncertainty in the temperature of the emitting region: If the temperature on day 199 would be $\sim 20,000$K (i.e., the interaction power is higher in the earlier phase), then the estimate goes up to $M({\rm H}) \sim 0.28 M_{\odot}$. Therefore, we conclude that the mass of emitting hydrogen is $M(H) \sim 0.2 M_{\odot}$. The lack of the evolution favors the scenario where the emitting region is the unshocked ejecta: If the H$\alpha$ emitting region is the RS, then the mass should have been increased by a factor of more than three from day 199 to 626. The mass we derived is therefore likely representative of the whole H-rich envelope. If we assume the composition we derived for SN 2013df is also applicable to SN 1993J, the estimate by \citet{patat1995} becomes $M({\rm H}) \sim 0.14 M_{\odot}$. Therefore, the mass of the envelope hydrogen is similar between SN 2013df and 1993J, with SN 2013df probably larger by a factor of $\sim 1.4$. By combining H and He, we derive the total mass of the emitting region as $\sim 0.5 M_{\odot}$. 

From the profile of H$\alpha$ emission, we estimate that the expansion velocity of the shell decreases following $V_{\rm sh} \propto t^{\beta}$ where $\beta \sim -0.04$ to $\sim -0.08$ (Figure 4d). The self-similar solution \citep{chevalier1982a} predicts $\beta = (3-s) / (m-s)$, where $m$ and $s$ are power law indexes in the density distribution as a function of velocity or radius for the SN ejecta and the CSM, respectively. Adopting $s = 2$ as expected for the steady-state mass loss which is (roughly) applicable to SNe 1993J \citep{fransson1989} and 2011dh \citep{maeda2014a}, we derive $m \sim 15 - 27$ which takes into account the uncertainty in measuring the deceleration of the shell velocity. While measuring the exact value of $m$ is difficult due to the sensitive dependence of $m$ to $\beta$, the outer ejecta structure of SN 2013df seems to be steep. This steep density gradient is difficult to reconcile with a standard model of the SN dynamics which predict $m \sim 7 - 12$ \citep{chevalier2006}, but is similar to what have been derived for SNe 1993J \citep{fransson1996} and 2011dh \citep{maeda2014a}. This indicates that our understanding of the outermost density structure in the H-rich envelope of the SN IIb progenitors is still missing \citep[see e.g., ][]{nakar2014}, and the fact that the steep density gradient is generally found in the well-studied SNe IIb should provide a hint in understanding the still unresolved final evolution of the progenitors toward SNe IIb. 

Applying the self-similar solution of \citet{chevalier1982a} for the velocity at the CD, specifically using eq. 5 of \citet{patat1995}, we can derive the CSM density, i.e., $A_{*}$ as defined by $\rho_{\rm CSM} = 5 \times 10^{11} A_{*} r^{-2}$ (i.e., $A_{*} = 100$ for the mass loss rate of $\sim 10^{-5} M_{\odot}$ yr$^{-1}$ and the wind velocity of $\sim 10$ km s$^{-1}$). For a range of $m \sim 15 - 27$, $M_{\rm ej} = 2 M_{\odot}$ \citep{morales2014}, and the characteristic ejecta velocity of $10,000$ km s$^{-1}$, we derive $A_{*}  \sim 110$ (for $m = 27$) and $A_{*} \sim 430$ (for $m = 15$) for SN 2013df. Therefore, we conclude that the CSM around SN 2013df is dense with $A_{*} \sim 110 - 430$. If this CSM density is converted to the mass loss rate under the assumption of a steady-state mass loss \citep[which roughly applies to SNe 1993J and 2011dh: ][]{maeda2014a} and the wind velocity of $\sim 20$ km s$^{-1}$, the mass loss rate in the final $\sim 800$ years before the SN explosion is $(5.4 \pm 3.2) \times 10^{-5} M_{\odot}$ yr$^{-1}$.  Throughout the paper, we adopt $20$ km s$^{-1}$ as our fiducial value for the SN IIb progenitor wind velocity; this is at the low end as inferred for YSGs \citep[$20 - 100$ km s$^{-1}$; ][]{smith2014}, and the wind velocity of $\sim 10 - 20$ km s$^{-1}$ have been assumed for the progenitors of SNe 1993J and 2011dh \citep{fransson1996,maeda2014a}. We note that in any case the uncertainty in the wind velocity by a factor of a few would not affect the conclusions in this paper. Moreover, the derived mass loss rate is simply scaled linearly with the assumed wind velocity. 

The derived CSM density is also largely consistent with X-ray properties of SN 2013df. SN 2013df was brighter than SN 2011dh in X-rays by a factor of $\sim 10$ \citep{li2013}. Figure 9 of \citet{maeda2014a}, taking into account of the cooling and absorption in the RS region, shows that the predicted X-ray luminosity is roughly consistent with that of SN 2013df for $A_{*} \sim 200 - 400$. Indeed, from a detailed analysis of the radio and X-ray properties of SN 2013df, \citet{kamble2015} independently derived the CSM density which is roughly in agreement with the estimate mentioned here. 

The mass in the RS region (i.e., the mass in the swept-up ejecta) can be estimated for given $A_{*}$ and $m$ (and $s=2$). \citet{chevalier1982b} showed the scaling relation as $M_{\rm RS} / M_{\rm FS} = (m-4)/2$, where $M_{\rm RS}$ and $M_{\rm FS}$ are the masses in the RS and FS regions. This formula overestimates the value found in a more detailed self-similar analysis of \citet{chevalier1982a} by $\sim 30$\%. Using these relations, we estimate $M_{\rm RS} \sim 0.28 M_{\odot}$ for $m = 15$ and $M_{\rm RS} \sim 0.15 M_{\odot}$ for $m = 27$. Note that $M_{\rm RS}$ is not sensitive to the exact value of $m$. Taking into account this uncertainty, $M_{\rm RS}$ does not seem to be sufficient to explain the mass of the H$\alpha$-emitting region ($\sim 0.5 M_{\odot}$), adding another support for the interpretation that H$\alpha$ is produced by recombinations within the unshocked ejecta. The temperature behind the RS is probably too high to contribute to the H$\alpha$ \citep[e.g., $1-3$ keV derived for SN 2011dh;][]{maeda2014a}, and then we derive the envelope mass as $\sim 0.7 M_{\odot}$ by adding this mass to that of the H$\alpha$ emitting region as derived above. Note that this estimate is fully independent from LC models using the early-phase LC at the first few days which is yet to be fully developed \citep[e.g., ][]{nakar2014}, and our late-time observation provides a good calibration for such models. Our result indicates similar structures in the progenitor envelopes between SNe 2013df and 1993J -- the H-rich envelope mass was estimated to be $\sim 0.3 - 1 M_{\odot}$ through the LC models for SN 1993J. Note that the H-rich envelope masses in SNe 2013df and 1993J are substantially larger than that of SN 2011dh, which was constrained to be $\sim 0.1 M_{\odot}$ \citep{arcavi2011,bersten2012,maeda2014a}. This again points to the difference between SN 2013df (and SN 1993J) and SN 2011dh. 

\subsection{A Link between the Natures of Progenitors and Their Mass Losses}

A sample of well-studied SNe IIb to date, including the pre-SN progenitor (candidate) detection, prompt multi-wavelength follow up in a few days since the explosion, and a long term follow-up covering more than 500 days, composes of only three SNe IIb (1993J, 2011dh, and 2013df). However, the similarity and difference among the sample are striking. It has been shown that these three are similar in the SN ejecta properties as inferred by their maximum-light properties \citep{morales2014,vandyk2014}. The nature of the progenitor seems to be divided into two classes, a more extended progenitor (SNe 1993J and 2013df) and a less extended one (SN 2011dh), as inferred both by the progenitor (candidate) in the pre-SN images \citep{maund2004,vandyk2013,vandyk2014} and by the behavior in the LC in the first few days \citep{morales2014,vandyk2014}. This classification and diversity are further extended by adding UV properties to the list of observables \citep{benami2015}. We find that this classification further applies to the late-time behavior at $\gsim 600$ days since the explosion. While the sample is still limited, the similarities between SNe 1993J and 2013df, as well as the contrast to SN 2011dh, are so striking, and we speculate that these two classes might reflect distinctly different paths in the evolution toward an SN. This might be related to the `extended' and `compact' classification for SNe IIb as proposed by \citet{chevalier2010} mostly from their radio properties \citep[see also][]{kamble2015}. 

In Figure 6, we show comparison of SNe IIb in properties of the early-phase `cooling' emission / progenitor radius and the CSM density / the mass loss rate in the final centuries toward the SN explosion. The properties inferred for SNe Ib/c are also indicated, while for SNe Ib/c the derivation of the CSM density and thus the mass loss rates is not as robust as that for the well-studied SNe IIb \citep[see, e.g., ][for the uncertainty]{maeda2012a}. In converting the CSM densities to the mass loss rates, we assume the mass loss wind velocity of $20$ km s$^{-1}$ for SNe IIb and $1,000$ km s$^{-1}$ for SNe Ib/c. 

There is a correlation between the strength of the early-cooling luminosity and the CSM density as shown in Figure 6a. SNe IIb 2011dh and 2008ax have the CSM density at the high-end assumed for SNe Ib/c, both of which have been classified as `compact' SNe IIb \citep{soderberg2012}.\footnote{Given the confirmed YSG progenitor of SN 2011dh, more appropriate terminology would be `less extended'.}  SNe IIb 2013df and 1993J clearly have a larger CSM density than these SNe IIb by an order of magnitude, and than typical SNe Ib/c by two orders of magnitudes. SN 1993J reached a peak in the early-cooling emission, as seen in the pseudo-bolometric optical-NIR LC, at $\sim 3-4$ days after the explosion. This is due to the combination of decreasing bolometric luminosity and decreasing temperature \citep[e.g., ][]{nakar2014}. As the early-cooling luminosity, we adopt this luminosity for SN 1993J, and the pseudo-bolometric optical-NIR luminosity (or upper limit) at the same epoch for the other SNe IIb \citep[see, e.g., figure 6 of ][]{morales2014}. SNe Ib/c and `compact' SNe IIb do not show signatures of the early-cooling emission (except for SN 2011dh thanks to a deep and prompt follow-up), but deep limits have been obtained for some SNe Ib/c \citep[e.g.,][]{cao2013}. No progenitor candidates have been reported for SNe Ib/c except for a possible WR progenitor of iPTF13bvn \citep{cao2013}.\footnote{However, we note that it has been claimed by \citet{eldridge2015} that the detected progenitor candidate is more likely a binary system consisting of lower mass stars than claimed by \citet{cao2013}.} In any case, a giant progenitor has been rejected for some SNe Ib/c \citep[e.g., see][for a review]{smartt2009,eldridge2013}. As expected, the relation is also clearly seen if we use the progenitor radius instead of the early-phase cooling luminosity (Figure 6b). This finding is consistent with the correlation between the properties of early-cooling phase and UV properties around maximum-light if the latter is interpreted as a trace of the SN-CSM interaction \citep{benami2015}.

While there seems to be a well-defined relation between the progenitor radius and the CSM density, it becomes less obvious if the CSM density is converted to the mass loss rate assuming typical wind velocities for a WR star ($\sim 1,000$ km s$^{-1}$) for SNe Ib/c and a YSG star ($\sim 20$ km s$^{-1}$) for SNe IIb (see \S 4.1), as shown in Figure 6c. We show this for the purpose of demonstration since the mass loss rate is more directly linked to the progenitor evolution than the CSM density, while the CSM density is more directly linked to the observations; the inferred mass loss rates here suffer from the uncertainty by a factor of a few in the wind velocity. In any case, our argument below is based on the difference in the mass loss rate by an order of magnitude, and thus the uncertainty in the wind velocity would not affect our conclusions. For SNe Ib/c this is derived for the radio observation at $\lsim 100$ days after the explosion corresponding to the final $\lsim 10$ years of the mass loss toward the explosion, and for SNe IIb the derivation is based on the observations at $\sim 500$ days corresponding to $\sim 1000$ years before the explosion. The range of the mass loss rate inferred for SNe Ib/c seems to overlap with the range seen in SNe IIb. This mostly comes from the high wind velocity for a hypothesized WR progenitor of SNe Ib/c, and from some SNe Ib/c that show a relatively high CSM density \citep{wellons2012}. In any case, if we focus on the well-observed SNe IIb, there is a relation (within the small sample) between the progenitor radius and the mass loss rate, divided into SNe 2013df and 1993J on one side (more extended and a larger mass loss rate) and SN 2011dh on the other (less extended and a smaller mass loss rate). 

In sum, we suggest that there is a relation between the nature of the progenitor (i.e., in the radius and mass of the H-rich envelope) and the mass loss rate in the final centuries toward the explosion, for a giant progenitor of SNe IIb. To clarify if this is a generic feature among SNe IIb requires a larger sample, but we see this does apply to the well-observed three SNe IIb 1993J, 2011dh and 2013df: Progenitors with a more extended and massive H-rich envelope have larger mass loss rates in the final centuries toward the explosion. This is at odds with a naive expectation: If a progenitor had a larger mass loss rate, it would lose a larger amount of the H-rich envelope, contrary to what we observe. One possibility is that the relation reflects a possible causality between the two quantities: With such a high CSM density for SNe 2013df and 1993J ($A_{*} \sim 200$), the electron scattering opacity in the wind region before the explosion, if the CSM was fully ionized (while unlikely for a giant progenitor), is estimated to be an order of unity near the vicinity of the surface of a progenitor. If this is the case, the progenitor radius might have been falsely identified in the photosphere within the wind region for SNe 2013df and 1993J. However, this does not explain another observational indication that we also see the difference between the two classes in the mass of the H-rich envelope. While further investigation is necessary, we regard this possibility as unlikely. 

Therefore, we suggest that this unexpected relation between the progenitor radius and the mass loss rate reflects a real difference in the evolutionary paths to these `two classes' of SNe IIb. Given that the binary interaction is proposed to be a key for these SNe IIb \citep[e.g.,][]{benvenuto2013} and that the mass loss rate is affected substantially by the binary interaction \citep[e.g., ][for a review]{smith2014}, this suggests that there is a difference in the binary interaction history for these two classes, reflecting differences in the evolution of the binary separation and mass transfer history. The mass loss rate just before the explosion was found to be insufficient to produce a YSG progenitor for SN 2011dh \citep{maeda2014a} and a strong binary interaction phase well before the explosion (at least $\gsim 1,000$ years before the explosion) is necessary. On the other hand, the derived mass loss rates for SNe 2013df and 1993J are indeed enough to expel nearly all the H-rich envelope in the giant phase (if this mass loss rate would keep that high during a whole giant phase), while the binary interaction is probably required to drive such a large mass loss rate in the giant phase. Therefore, it seems that the timing of the binary interaction might be different between the two classes. As an emerging picture, we propose that SNe IIb with a more extended progenitor are explosions during a strong binary interaction phase (including any effects by a companion star such as irradiation), while those with a less extended progenitor have a time delay between the strong binary interaction phase and the explosion. A further investigation of a possible companion star for SN 2013df, whose nature should be related to the binary interaction history, will be an important step to solve this issue, as complemented by reported detections of possible companion candidates for SNe 1993J \citep{maund2004,fox2014} and 2011dh \citep{folatelli2014b}. 

\section{Concluding Remarks}
In this paper, we have reported the late-time evolution of Type IIb Supernova (SN IIb) 2013df, for which the progenitor candidate had been identified as a yellow supergiant. SN 2013df showed a dramatic change in its spectra at $\sim 1$ year. On day 626, it has completely lost resemblance to other SNe IIb/Ib/Ic, but has been dominated by broad and strong H$\alpha$ and He I emission lines arising from a thin shell expanding at the velocity of $\sim 10,000$ km s$^{-1}$. Indeed, the only previous example showing this characteristics was SN 1993J. Namely, SN 2013df turns out to be a twin of SN 1993J, not only in the early-phase properties but also in the late-time properties. This behavior is quite different from that seen in another well-observed SN IIb 2011dh, which also showed noticeable differences in the (very) early-phase properties and in the progenitor than SNe 2013df and 1993J. 

We have shown that this late-time evolution is fully explained by a change in the energy input from the $^{56}$Co decay to the interaction between the SN ejecta and dense CSM. We have derived some characteristic quantities in the SN ejecta and CSM under the interaction scenario, including the steep density gradient in the outermost SN ejecta, the mass and composition within the H-rich envelope as are consistent with models for SN 1993J (but the mass substantially larger than SN 2011dh), the CSM density, and the mass loss in the final centuries toward the explosion. 

SN 2013df on day 626 was found to be brighter than a canonical $^{56}$Co-powered model prediction at least by two magnitudes, as consistent with the SN-CSM interaction scenario. A pros is that we have a chance to observe the SN-CSM interaction for a long span to investigate the mass loss history further toward the past, but a con is that it will raise a hurdle for the confirmation of the progenitor disappearance and the detection of a possible companion. For the SN-CSM interaction study, we expect that the spectral features will further change so that [O III] 4959, 5007 will increase in strength as observed for SN 1993J \citep{matheson2000a} and also for middle-aged SNe IIn and IIp at $\gsim 10$ years after the explosion \citep{milisavljevic2012}. This transition will take place when the RS reaches to the bottom of the H-rich layer \citep[e.g., ][]{milisavljevic2012}, and therefore observing this transition will tell us further on the similarity and difference between SNe 2013df and 1993J. On the progenitor, we predict that, taking into account the interaction power, the SN will keep brighter in the $V$ and $R$-bands than the progenitor candidate until early 2016, or even so until late 2017. A continuous follow-up of this SN is therefore very interesting in many aspects. 

The mass loss rate corresponding to the derived CSM density is $(5.4 \pm 3.2) \times 10^{-5} M_{\odot}$ yr$^{-1}$ (for the wind velocity of $20$ km s$^{-1}$), which is similar to SN 1993J but larger than SN IIb 2011dh by more than an order of magnitude. The striking similarity between SNe 2013df and 1993J in the progenitor (candidate) and the CSM environment, and the contrast in these natures as compared to SN 2011dh, infer that these is a link between the natures of the progenitor and the mass loss, where the SNe IIb with a more extended progenitor have experienced a much stronger mass loss than those with a more compact progenitor in the final centuries toward the explosion. We note that a similar conclusion has been independently reached by a detailed modeling of radio and X-ray properties \citep{kamble2015}, strengthening the case. Also, there is accumulating evidence that a binary interaction is a key in the whole SN IIb population including the present study, and so far there is no single SN IIb for which a single star evolution is favored \citep[see, e.g., ][]{kuncarayakti2015}. The link suggests that there might be two classes in SNe IIb in terms of characteristic binary evolution schemes: it might indicate that the timing between the strong binary interaction and the explosion might be different between the two classes, which might be a key to understand a unified scheme of the yet-unresolved evolution toward SNe IIb/Ib/Ic in general.

\acknowledgements 
The authors thank the staff at the Subaru Telescope for their excellent support in the observations. The authors thank Takaya Nozawa for his constructive comments on the manuscript. The work by K.M. is supported by JSPS Grant-in-Aid for Scientific Research (No. 26800100) and by WPI Initiative, MEXT, Japan. Support for H.K. is provided by the Ministry of Economy, Development, and Tourism's Millennium Science Initiative through grant IC120009, awarded to The Millennium Institute of Astrophysics, MAS. H.K. acknowledges support by CONICYT through FONDECYT grant 3140563. The work by M.T. is supported by JSPS Grant-in-Aid for Scientific Research (15H02075, 15H00788). 


\clearpage
\begin{figure*}
\begin{center}
        \begin{minipage}[]{0.7\textwidth}
                \epsscale{1.0}
                \plotone{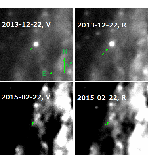}
        \end{minipage}\end{center}
\caption
{$V$ and $R$-band images of SN 2013df at 199 and 626 days after the explosion. 
\label{fig1}}
\end{figure*}

\clearpage
\begin{figure*}
\begin{center}
        \begin{minipage}[]{0.9\textwidth}
                \epsscale{0.7}
                \plotone{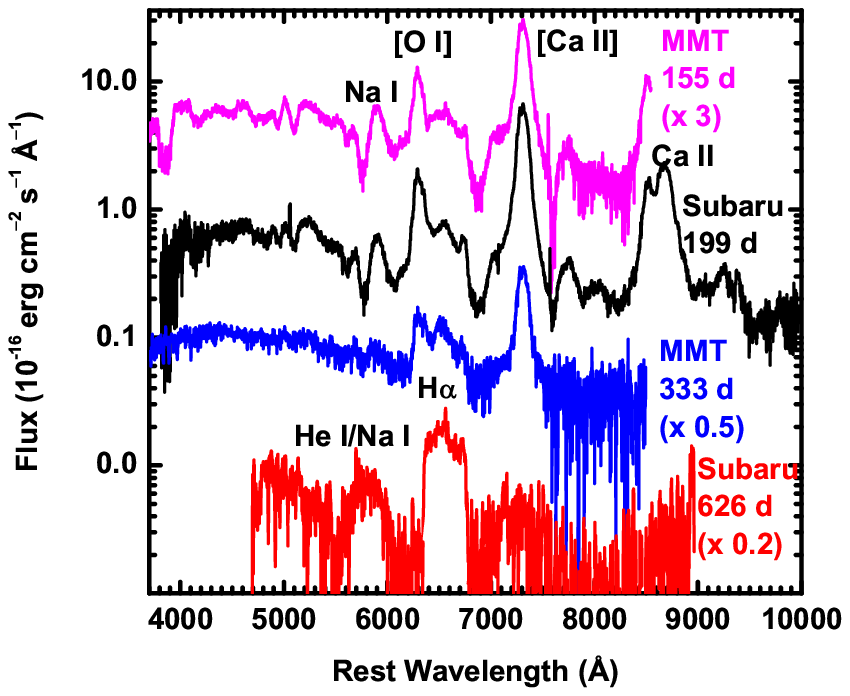}
        \end{minipage}\end{center}
\caption
{Time sequence of late-time spectra of SN 2013df from day 155 to day 626 after the explosion. For demonstration, the fluxes in the spectra except for the one on day 199 are multiplied by the amounts indicated in the figure. The extinction is corrected for with $E(B-V) = 0.1$ mag \citep{morales2014,vandyk2014}, assuming $R_{V} = 3.1$. 
\label{fig2}}
\end{figure*}

\clearpage
\begin{figure*}
\begin{center}
        \begin{minipage}[]{0.45\textwidth}
                \epsscale{1.0}
                \plotone{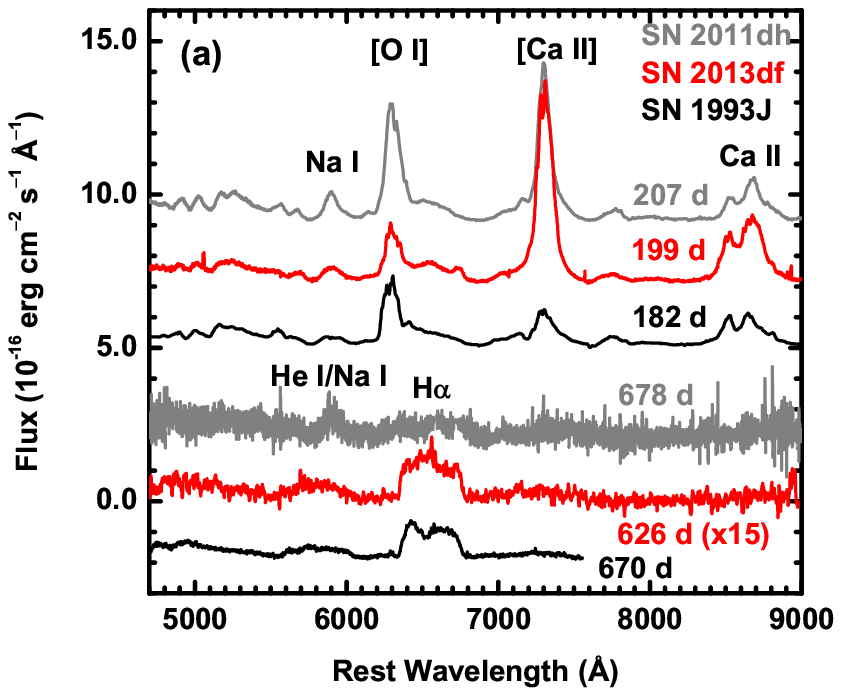}
        \end{minipage}
         \begin{minipage}[]{0.45\textwidth}
                \epsscale{1.0}
                \plotone{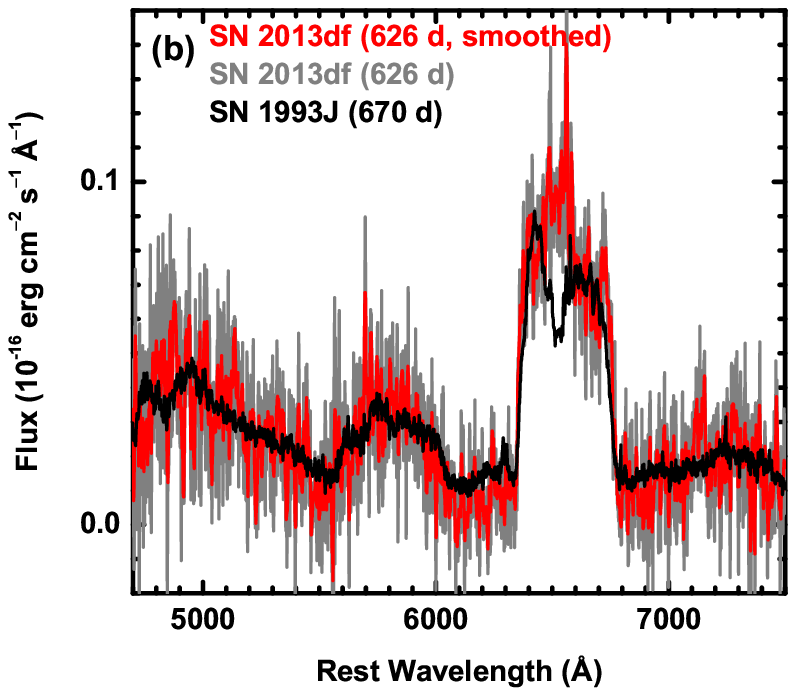}
        \end{minipage}
\end{center}
\caption
{(a) Late-time spectra of SN 2013df (red) as compared to those of SNe 1993J \citep[black:][]{matheson2000a,matheson2000b} and 2011dh \citep[gray:][]{shivvers2013} at the similar epochs. The extinction is corrected for with $E(B-V) = 0.1$ \citep{morales2014,vandyk2014}, $0.17$ \citep{ergon2014a}, and $0.07$ \citep{ergon2014a} mag for SNe 2013df, 1993J, and 2011dh, respectively, assuming $R_{V} = 3.1$. The flux scale on the vertical axis is for SN 2013df at day 119; the flux at day 199 is added by a constant, and that at day 626 is multiplied by a factor of 15 for presentation. The fluxes of the comparison spectra are either multiplied or added by artificial amounts for presentation. The spectrum of SN 2013df at day 626 is smoothed with a 5 pixel boxcar filter. (b) A close comparison between the spectra of SN 2013df at day 636 (non-smoothed spectrum shown by gray, and the smoothed one shown by red) and that of SN 1993J at day 670 (black). 
\label{fig3}}
\end{figure*}

\clearpage
\begin{figure*}
\begin{center}
        \begin{minipage}[]{0.45\textwidth}
                \epsscale{1.0}
                \plotone{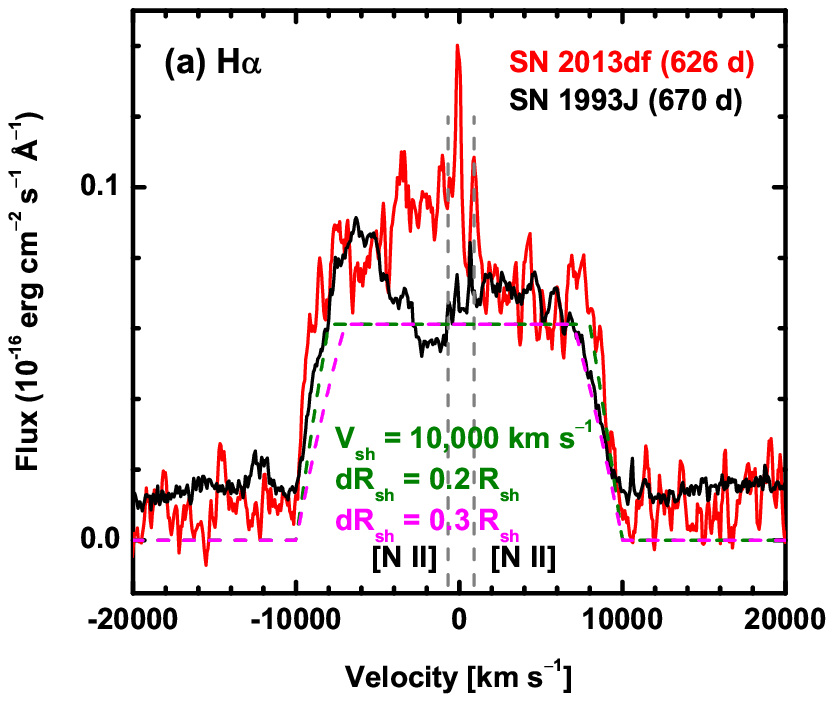}
        \end{minipage}
         \begin{minipage}[]{0.45\textwidth}
                \epsscale{1.0}
                \plotone{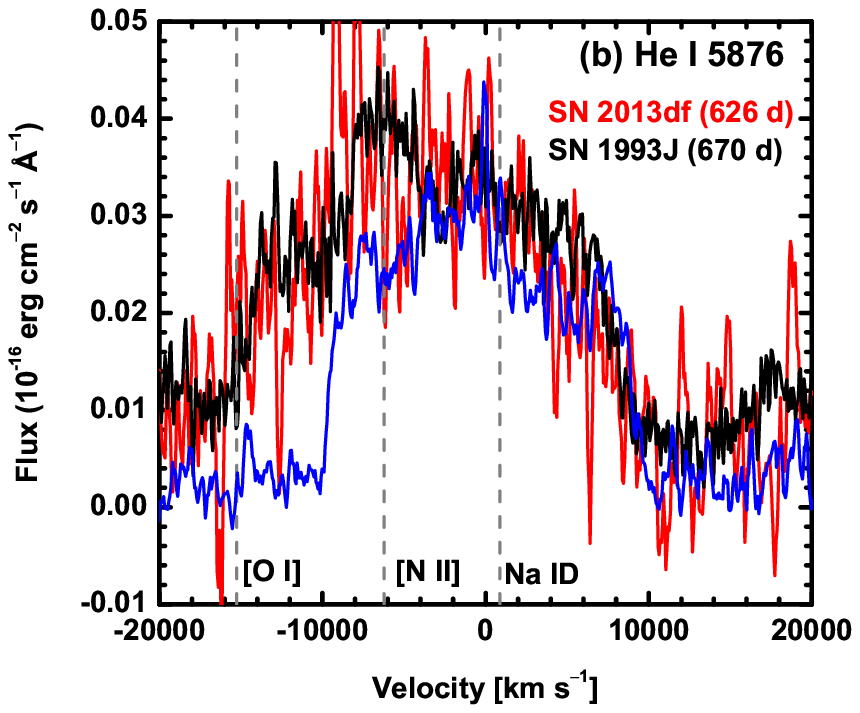}
        \end{minipage}
         \begin{minipage}[]{0.45\textwidth}
                \epsscale{1.0}
                \plotone{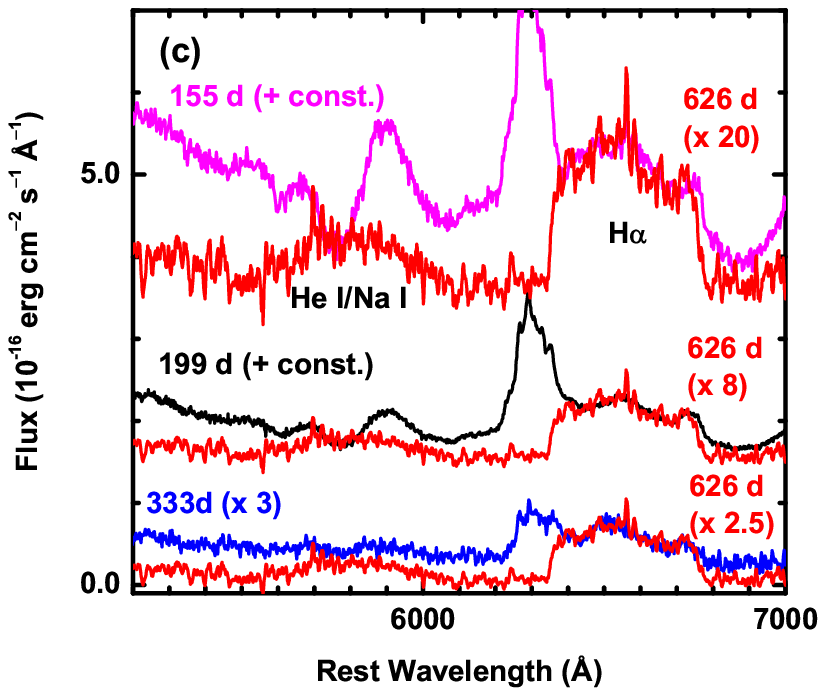}
        \end{minipage}
         \begin{minipage}[]{0.45\textwidth}
                \epsscale{1.0}
                \plotone{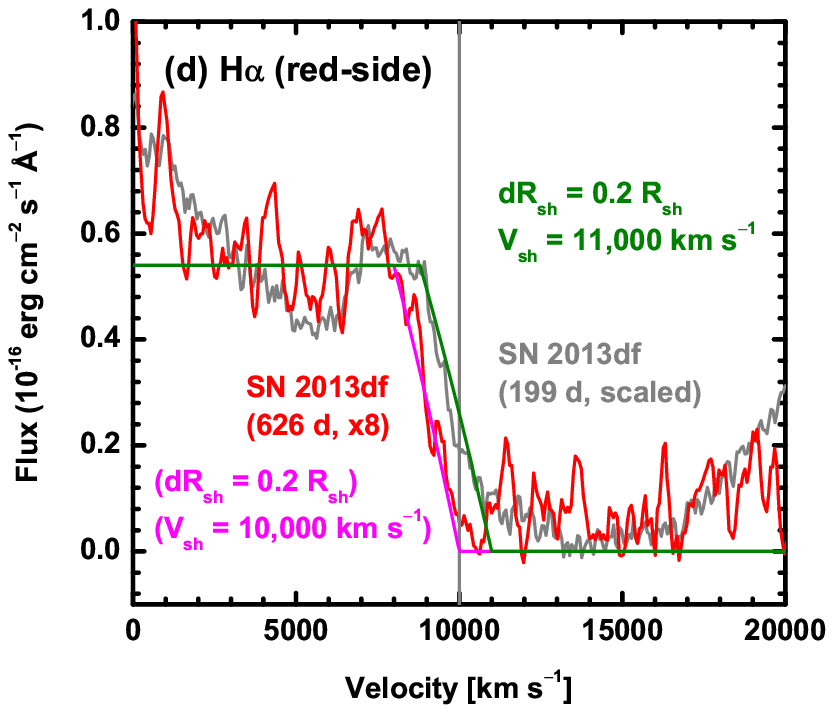}
        \end{minipage}
\end{center}
\caption
{(a) H$\alpha$ and (b) He I 5876 (or Na ID) in the spectrum of SN 2013df at day 626 (red) as compared to those of SN 1993J (black). In (a), we plot expected profiles computed for the emission from a thin shell with the outer edge velocity of $V_{\rm sh } = 10,000$ km s$^{-1}$ and the shell width ($\Delta R_{\rm sh}/R_{\rm sh}$) of $0.2$ (green-dashed) and 0.3 (magenta-dashed): The thin shell predicts a flat profile below the wavelength corresponding to $V_{\rm sh} - \Delta V_{\rm sh}$ while a parabola between $V_{\rm sh} - \Delta V_{\rm sh}$ and $V_{\rm sh}$, where $\Delta V_{\rm sh} = \Delta R_{\rm sh}/t$. The former ($\Delta R_{\rm sh}/R_{\rm sh} = 0.2$) provides a reasonable fit to the observed profile, while the latter ($\Delta R_{\rm sh}/R_{\rm sh} = 0.3$) is clearly rejected. In (b), H$\alpha$ of SN 2013df at day 626, as divided by a factor of $3.2$, is shown (blue). The rest wavelengths of possible contaminating lines are denoted by dashed vertical lines ([O I] 5577, [N II] 5754, [N II] 6548, [N II] 6583). (c) Evolution of SN 2013df from day 155 to 626, where the spectrum on day 626 multiplied by a factor of 20, 8, and 2.5 is compared to the spectra on day 155, 199, and 333, respectively. (d) An expanded view on the evolution of the H$\alpha$ profile from day 199 (gray) to day 626 (red). The thin shell models that provide a reasonable fit to each of the spectra are shown for the one at day 199 (green) and day 626 (magenta), which quantify the velocity decrease between the two epochs. 
\label{fig4}}
\end{figure*}

\clearpage
\begin{figure*}
\begin{center}
        \begin{minipage}[]{0.45\textwidth}
                \epsscale{1.0}
                \plotone{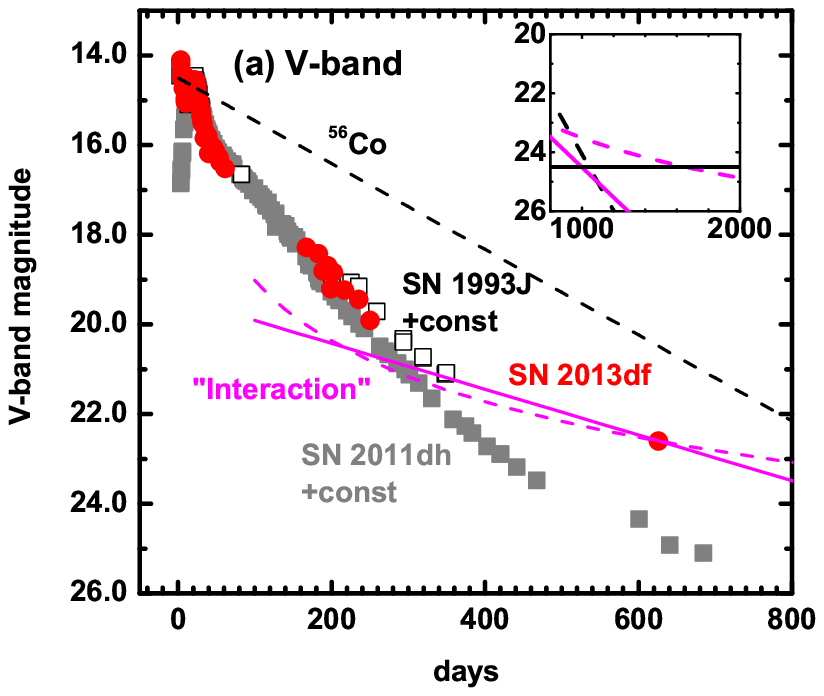}
        \end{minipage}
         \begin{minipage}[]{0.45\textwidth}
                \epsscale{1.0}
                \plotone{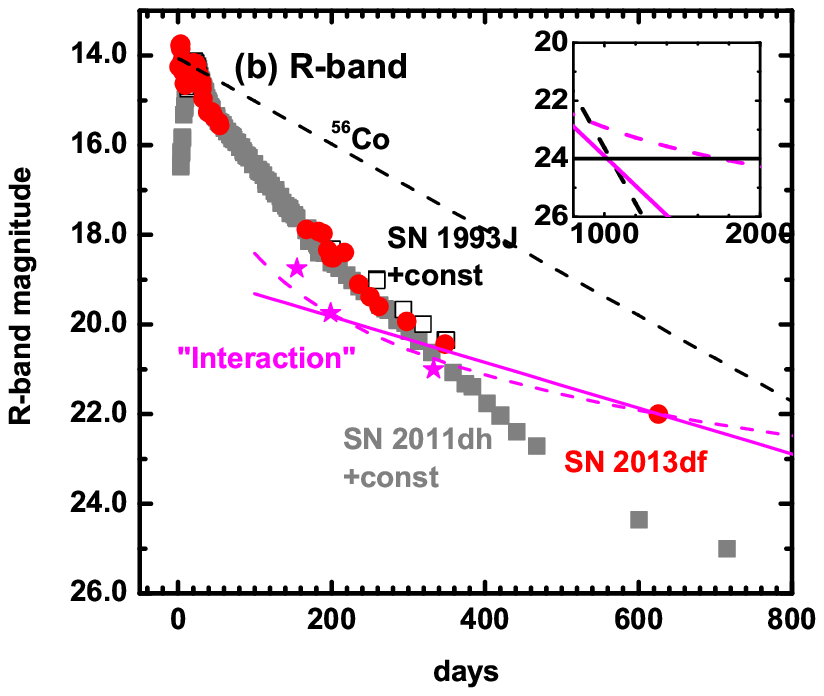}
        \end{minipage}
\end{center}
\caption
{$V$ and $R$-band light curves of SN 2013df \citep[red points:][and this paper]{morales2014} as compared to those of SNe 1993J \citep[black open squares:][]{barbon1995} and 2011dh \citep[gray filled squares:][]{ergon2014b}. The magnitudes of SNe 1993J and 2011dh are artificially shifted to match to the magnitude scale of SN 2013df. The black-dashed line shows the $^{56}$Co decay line (for a full deposition). In the $R$-band, the contribution of the interaction power and its evolution (stars) are estimated by the H$\alpha$ flux for epochs where the spectra are available (Figure 4c). The same behavior is assumed for the $V$-band. A fit to the interaction light curves is shown for a functional form of exponential (magenta-solid) and power-law (magenta-dashed). The $^{56}$Co decay power (represented by the late-time evolution of SN 2011dh) is insufficient to explain the brightness of SN 2013df on day 626. Also, the late-time LC evolution of SN 2013df is much flatter than the prediction from the $^{56}$Co-decay input, requiring another power source. The inserted panels show the $^{56}$Co decay (full deposition) and extrapolation of the interaction light curves in the later epochs, together with the magnitude of the candidate progenitor. 
\label{fig5}}
\end{figure*}

\clearpage
\begin{figure*}
\begin{center}
        \begin{minipage}[]{0.32\textwidth}
                \epsscale{1.0}
                \plotone{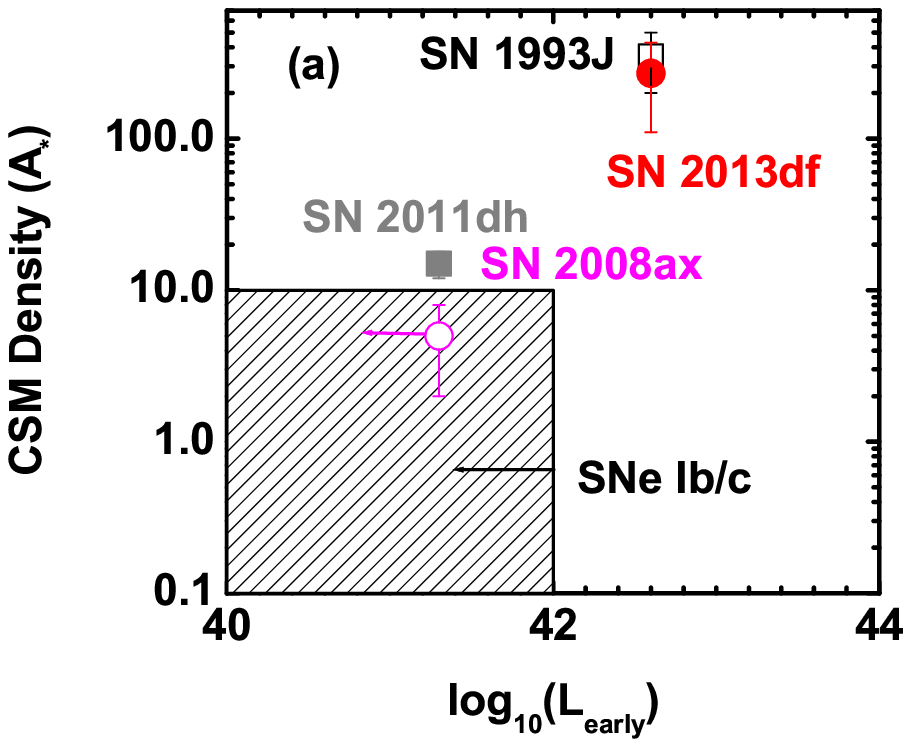}
        \end{minipage}
         \begin{minipage}[]{0.32\textwidth}
                \epsscale{0.92}
                \plotone{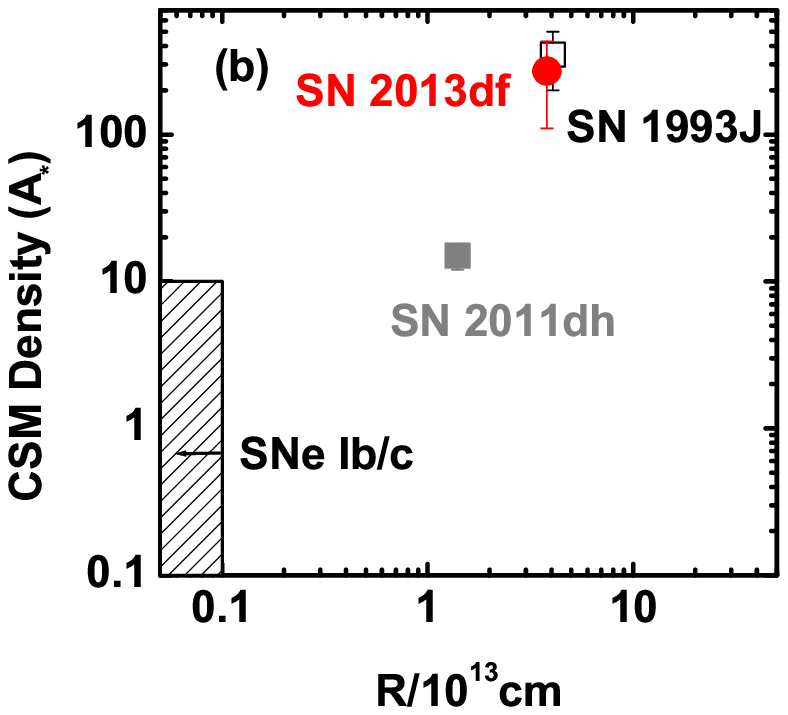}
        \end{minipage}
         \begin{minipage}[]{0.32\textwidth}
                \epsscale{1.0}
                \plotone{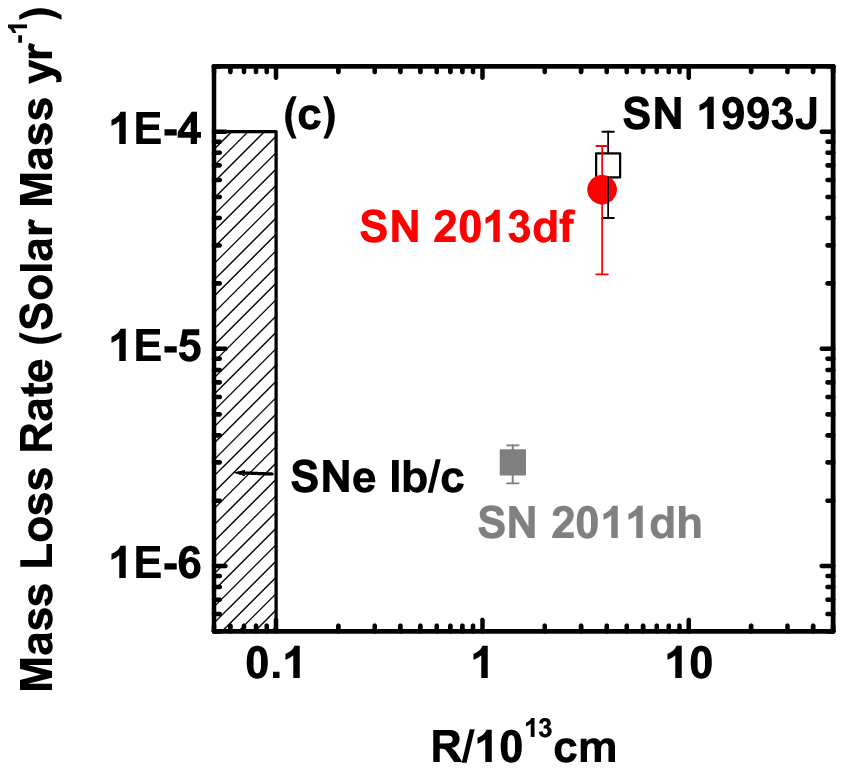}
        \end{minipage}
\end{center}
\caption
{The relations (a) between the pseudo-bolometric optical-NIR luminosity in the early `cooling' phase \citep[upper limits for SN IIb 2008ax and for SN Ib iPTF13bvn: ][]{pastorello2008,cao2013} and the CSM density, (b) between the progenitor radius and the CSM density, and (c) between the progenitor radius and the mass loss rate. The CSM density is described by $A_{*}$ as defined by $\rho_{\rm CSM} = 5 \times 10^{11} A_{*} r^{-2}$. To convert the CSM density to the mass loss rate, we have assumed the wind velocity of $20$ km s$^{-1}$ for SNe IIb and $1000$ km s$^{-1}$ for SNe Ib/c. The solid estimate of the CSM density has been provided for SNe 1993J \citep{fransson1996} and 2011dh \citep{maeda2014a} using the combined information from radio and X-ray data. For SN 2008ax, we convert the value estimated with radio data \citep{chevalier2010} to fit to the non-thermal acceleration efficiency derived for SN 2011dh \citep{maeda2014a}. The typical CSM densities assumed for SNe Ib/c are indicated by the shaded region \citep{chevalier2006}, while it may contain a systematic uncertainty of up to an order of magnitude \citep{maeda2012a}. The progenitor radii are plotted in (b, c) for those with the progenitor detection \citep[SNe 1993J and 2011dh:][]{maund2004,vandyk2013} and with the progenitor candidate detection \citep[SN 2013df:][]{vandyk2014}. For SNe Ib/c, upper limits on the magnitudes of some SN Ib/c sites in pre-SN images are consistent with a WR star, including a possible progenitor detection of iPTF13bvn, i.e., $\lsim 10R_{\odot}$ as inferred by the shaded area. 
\label{fig6}}
\end{figure*}

\end{document}